\documentclass[showpacs,prb,floatfix,superscriptaddress,twocolumn,amsmath]{revtex4-1}
\usepackage[T1]{fontenc}
\usepackage[latin1]{inputenc}
\usepackage[english]{babel}
\usepackage{epsfig}
\usepackage{color}
\usepackage{amssymb}  
\usepackage{amsmath}  
\usepackage{latexsym} 
\usepackage{amsthm}   
\newcommand{\be}{\begin{equation}}
\newcommand{\ee}{\end{equation}}
\newcommand{\bea}{\begin{eqnarray}}
\newcommand{\eea}{\end{eqnarray}}
\newcommand{\ba}{\begin{array}}
\newcommand{\ea}{\end{array}}

\newcommand{\gr}[1]{\mathbf{#1}}

\newcommand{\kvec}{{\bf k}}
\newcommand{\qvec}{{\bf q}}

\begin{document}\title{Signatures of nematic quantum critical fluctuations in the Raman spectra of lightly doped
cuprates}           

\author{S. Caprara}
\affiliation{Dipartimento di Fisica, Universit\`a di Roma Sapienza, 
Piazzale Aldo Moro 5, I-00185 Roma, Italy}
\affiliation{Istituto dei Sistemi Complessi CNR and CNISM Unit\`a di Roma Sapienza}
\author{M. Colonna}
\affiliation{Dipartimento di Fisica, Universit\`a di Roma Sapienza, 
Piazzale Aldo Moro 5, I-00185 Roma, Italy}
\author{C. Di Castro}
\affiliation{Dipartimento di Fisica, Universit\`a di Roma Sapienza, 
Piazzale Aldo Moro 5, I-00185 Roma, Italy}
\affiliation{Istituto dei Sistemi Complessi CNR and CNISM Unit\`a di Roma Sapienza}
\author{R. Hackl}
\affiliation{Walther Meissner Institut, Bayerische Akademie der Wissenschaften, 85748 Garching, Germany}
\author{B. Muschler}
\affiliation{Walther Meissner Institut, Bayerische Akademie der Wissenschaften, 85748 Garching, Germany}
\author{L. Tassini}
\affiliation{Walther Meissner Institut, Bayerische Akademie der Wissenschaften, 85748 Garching, Germany}
\author{M. Grilli}
\affiliation{Dipartimento di Fisica, Universit\`a di Roma Sapienza, 
Piazzale Aldo Moro 5, I-00185 Roma, Italy}
\affiliation{Istituto dei Sistemi Complessi CNR and CNISM Unit\`a di Roma Sapienza}

\begin{abstract}
We consider the lightly doped cuprates Y$_{0.97}$Ca$_{0.03}$BaCuO$_{6.05}$ and
La$_{2-x}$Sr$_x$CuO$_4$ (with $x=0.02$,\,0.04), where the presence of a fluctuating nematic state has often 
been proposed as a precursor of the stripe (or, more generically, charge-density wave) phase, which sets in 
at higher doping. We phenomenologically assume a quantum critical character for the longitudinal and transverse 
nematic, and for the charge-ordering fluctuations, and investigate the effects of these fluctuations in Raman 
spectra. We find that the longitudinal nematic fluctuations peaked at zero transferred momentum account well 
for the anomalous Raman absorption observed in these systems in the $B_{2g}$ channel, while the absence of such 
effect in the $B_{1g}$ channel may be due to the overall suppression of Raman response at low frequencies, associated 
with the pseudogap. While in Y$_{0.97}$Ca$_{0.03}$BaCuO$_{6.05}$ the low-frequency lineshape is fully accounted 
by longitudinal nematic collective modes alone, in La$_{2-x}$Sr$_x$CuO$_4$ also charge-ordering modes with 
finite characteristic wavevector are needed to reproduce the shoulders observed in the Raman response. This 
different involvement of the nearly critical modes in the two materials suggests a different evolution of the 
nematic state at very low doping into the nearly charge-ordered state at higher doping.
\end{abstract} 

\date{\today} 
\pacs{74.72.-h, 74.25.nd, 75.25.Dk, 74.40.Kb}
\maketitle

\section{Introduction}
\label{intro}
Growing experimental and theoretical evidence indicates that (stripe-like) charge ordering (CO) 
[\onlinecite{kivelson,FPS,sachdev2013}], possibly related to a hidden charge-density-wave quantum critical point 
near optimal doping [\onlinecite{CDG,andergassen,reviewQCP1,reviewQCP2,chubukov2014}], plays a role in 
determining the unconventional properties of superconducting cuprates. Charge ordered textures 
were assessed by neutron scattering experiments in La cuprates, codoped with Nd [\onlinecite{tra95,tra96,tra97}], 
Ba [\onlinecite{fujita02}], or Eu [\onlinecite{klauss00}], and confirmed also by soft resonant x-ray scattering 
[\onlinecite{abb05},\onlinecite{fink09}]. The occurrence of stripe-like charge- and spin-density waves in other 
cuprates is supported by the similarities of the non-codoped and codoped La cuprates in the 
spin channel, e.g., the doping dependence of the low-energy incommensurability [\onlinecite{yam98}], and the 
high-energy magnon spectra in La$_{2-x}$Ba$_{x}$CuO$_4$ [\onlinecite{tra04}], La$_{2-x}$Sr$_{x}$CuO$_4$ 
(LSCO) [\onlinecite{chr04}], and YBaCuO$_{6+p}$ [\onlinecite{hay04},\onlinecite{hinkov08}]. These features are 
well described in terms of striped ground states [\onlinecite{lor02,sei05,sei06}]. CO in cuprates, possibly 
with fluctuating character, was also confirmed by EXAFS [\onlinecite{saini}], NMR experiments 
[\onlinecite{julien1},\onlinecite{julien2}], scanning tunneling spectroscopy 
[\onlinecite{kapitulnik,yazdani,davis}], and resonant x-ray measurements 
[\onlinecite{ghiringhelli,chang2,damascelli,blancocanosa}]. A recent theoretical analysis of Raman spectra in LSCO 
[\onlinecite{CDMPHLEKAG}] showed that nearly critical spin and charge fluctuations coexist at intermediate and high 
doping. This coexistence also accounts [\onlinecite{mazza}] for the specific momentum, energy and doping dependence of 
the single-particle anomalies, the so-called kinks and waterfalls, observed in photoemission spectra 
[\onlinecite{reviewARPES}].
 
The above facts, support the occurrence and relevance of (fluctuating) stripes in cuprates and raise the 
question about their precursors at very low doping [\onlinecite{millis2010}]. The experimental evidence of rotational symmetry 
breaking [\onlinecite{hinkov08,haug,daou,mesaros2011,lawler2010}] points towards nematic order, although it is not 
yet clear whether this order arises from a melted stripe state [\onlinecite{KFE}], from incipient unidirectional 
fluctuating stripes [\onlinecite{vojta}], or from an unrelated $d$-wave-type nematic order which preserves 
translational symmetry [\onlinecite{sun}]. On the theoretical side, it was recently proposed that a ferronematic state
occurs at very low doping, formed by stripe segments without positional order [\onlinecite{SCDGL}]. 
These segments are oriented because they sustain a vortex and an antivortex of the antiferromagnetic order at 
their extremes, and break rotational and inversion symmetry. This phase has no order in the charge sector, but 
induces incommensurate peaks in excellent agreement with experiments in LSCO [\onlinecite{wakimoto}]. Recent Monte 
Carlo calculations [\onlinecite{capati}] showed that, lowering the temperature, the ferronematic state turns into 
a ferrosmectic state, where the segments have a typical lateral distance $\ell_c$, corresponding to CO with 
a characteristic wavevector $\qvec_c$ (with $|\qvec_c|\sim 1/\ell_c$). The segments thus appear as the natural 
precursors of stripes.

It is therefore important to assess nematic order in cuprates. The aim of the present work is to identify the 
signatures of nematic fluctuations in Raman scattering. This is a bulk (nearly surface-insensitive) probe and 
measures a response function analogous to that of optical conductivity [\onlinecite{Shastry:1990}]. However, while 
the latter averages over the Brillouin zone (BZ), different polarizations of the incoming and outgoing photons 
weight different parts of the BZ in Raman scattering [\onlinecite{Devereaux:2007}], introducing specific form factors. 
It turns out that the so-called $B_{1g}$ and $B_{2g}$ channels are the most relevant to extract the contributions 
of collective modes (CMs) in cuprates. We already investigated how these form factors can be exploited to identify 
the contributions of different (e.g., charge and spin) critical CMs, based on their different finite wavevectors 
[\onlinecite{CDMPHLEKAG},\onlinecite{suppa},\onlinecite{CDEG,sces08,schachinger}]. 
There are two classes of CM contributions. In one class, the 
CMs dress the fermion quasiparticles, introducing self-energy and vertex corrections, which affect the Raman spectra 
up to substantial fractions of eV [\onlinecite{CDMPHLEKAG},\onlinecite{CDEG},\onlinecite{sces08}]. In the other class, 
the excitation of pairs of CMs [\onlinecite{suppa}], reminiscent of the Aslamazov-Larkin (AL) paraconductive 
fluctuations near the metal-superconductor transition (see Fig.\,\ref{raman-diagrams}), 
affects mainly the low frequency part of the spectrum and produces an anomalous 
absorption up to few hundreds of cm$^{-1}$, as indeed observed, e.g., in LSCO [\onlinecite{tassini}]. The analysis 
for LSCO [\onlinecite{suppa}] was based on CO CMs with finite wavevector ${\bf q}_c$, while the 
role of spin CMs was ruled out by symmetry arguments. At moderate doping, the value 
${\bf q}_c \approx (\pm \pi/2,0), (0,\pm \pi/2)$ was deduced from inelastic neutron scattering as the double of 
the wavevector of spin incommensuration [\onlinecite{yam98}], within the stripe scheme (we use hereafter a square 
unit cell on the CuO$_2$ planes, with lattice spacing $a=1$). By symmetry arguments, and in 
agreement with experiments, fluctuations with such ${\bf q}_c$ give rise to an anomalous absorption 
in the $B_{1g}$ channel only. A rotated ${\bf q}_c \approx  2\pi(\pm 2x,\pm 2x)$ occurs for $x<0.05$, 
[\onlinecite{wakimoto}], making the anomalous Raman absorption show in the $B_{2g}$ channel only, consistent with 
the experiments. However, a similar anomalous absorption in the $B_{2g}$ 
channel is observed in Y$_{1-y}$Ca$_y$Ba$_2$Cu$_3$O$_{6+x}$ (YBCO) for doping $p(x,y)$ between 0.01 and 0.06 
[\onlinecite{tassini2008}]. Recent measurements [\onlinecite{haug}] do not support the rotation of the spin 
modulation vector in YBCO, at least down to $p=0.05$, and the extrapolation of the available data indicates that 
spin incommensuration disappears for $p\approx 0.02-0.03$, while CO seems to disappear for 
$p<0.08$ [\onlinecite{blancocanosa}]. Thus, if only CO fluctuations were to play a role, the anomalous 
peak observed in YBCO in the  $B_{2g}$ channel would be unexplained. Furthermore, CO CMs yield in LSCO spectra 
that are fully satisfactory at $x=0.1$, but less convincing at $x=0.02$, where the experimental lineshape 
seems to have a composite character, with a main peak accompanied by a shoulder at slightly higher frequencies. 
This suggests the presence of two CMs contributing to the anomalous absorption in the $B_{2g}$ channel at low doping 
in LSCO and raises the question about the nature of the additional CM. The uncertain situation with YBCO and 
the compositeness of the LSCO spectra call for a critical revision of the results of Ref.\,[\onlinecite{suppa}].

The above mentioned evidences for nematic order make it natural to inquire whether the 
anomalous Raman absorption observed in underdoped cuprates might be due to nematic  
fluctuations (not considered in Ref.\,[\onlinecite{suppa}]), possibly mixed with CO 
fluctuations (in LSCO). Therefore, within the same formal scheme of Ref.\,[\onlinecite{suppa}], we include here
the contribution of nematic fluctuations. We find indeed that at low doping the observed anomalous absorption can be 
due to the excitation of long-wavelength overdamped nematic fluctuations with longitudinal character, whose strong 
dynamics is apt to reproduce the observed lineshape. While in strongly underdoped YBCO this is enough, in LSCO a 
secondary CM with finite characteristic wavevector, which we identify with the CO CM, is needed to better represent 
the lineshape. The doping dependence of the lineshape in LSCO indicates that there is an evolution from 
a dominating NCM towards a major relevance of the CO CM, upon increasing doping.

The scheme of the paper is the following: In Sec.\,\ref{qpcm} we introduce a phenomenological model of 
fermion quasiparticles coupled to nearly critical CO CMs and NCMs in underdoped cuprates. Then, we proceed 
with the theoretical calculation of the Raman response due to these CMs (Secs.\,\ref{fCM}, \ref{floop}, and 
\ref{Rresp}). In Sec.\,\ref{results}, we compare the theoretical results with available Raman spectra for 
underdoped YBCO and LSCO. Sec.\,\ref{disconcl} contains our final remarks and conclusions. 
Appendix A contains some details about the calculations of the Feynman diagrams involved in the
anomalous Raman response. Details of the fitting procedure are found in Appendix B, while 
a discussion on the role of the pseudogap in the fermionic spectrum is found in Appendix C.

\section{The fermion-collective mode model and the Raman response}
\label{qpcm}
\subsection{The fermion-collective mode model}
\label{fCM}
We consider a phenomenological model where, similarly to the electron-phonon coupling, electrons are coupled to 
NCMs or CO CMs. This approach relies on the presence of fermion quasiparticles. This assumption, which is natural 
in the metallic phase of cuprates, is still justified in the strongly underdoped phase, where angle resolved 
photoemission [\onlinecite{yoshida,zhou}] and transport experiments [\onlinecite{ando,ando2}] highlight the presence of fermionic 
low-energy states (the so-called Fermi arcs) with a substantial mobility, indicating that fermion quasiparticles 
still survive in this ``difficult habitat''. Thus, we adopt the Hamiltonian
\begin{equation}
{\cal H}=\sum_{\kvec , \sigma}\xi_{\kvec}c^\dagger_{\kvec \sigma}c_{\kvec \sigma}
+\sum_{\kvec,\qvec,\sigma}\sum_\lambda g_\lambda(\kvec,\qvec)
c^\dagger_{\kvec +\qvec\sigma}c_{\kvec \sigma}
\Phi^\lambda_{-\qvec}
\label{hamiltonian}
\end{equation}
where $c^\dagger_{\kvec \sigma}$ ($c_{\kvec \sigma}$) creates (annihilates) a fermion quasiparticle with momentum 
$\kvec$ and spin projection $\sigma$, and $\xi_\kvec$ is the fermion dispersion on the CuO$_2$ planes 
of LSCO or YBCO (measured with respect to the chemical potential). Its specific form is rather immaterial for our 
analysis, once the generic shape of the Fermi surface of cuprates is taken into account. The index $\lambda$ labels 
transverse ($\lambda=t$) or longitudinal ($\lambda=\ell$) nematic fluctuations 
[\onlinecite{oganesyan},\onlinecite{garst}], and charge fluctuations ($\lambda=c$), represented by the boson 
fields $\Phi^\lambda$. The quasiparticles couple to NCMs via 
$g_{\lambda}(\kvec,\qvec)\equiv g_\lambda d^\lambda_{\kvec,\qvec}$, with 
$d^\ell_{\kvec,\qvec} = \cos(2\varphi_{\kvec,\qvec})$ and $d^t_{\kvec,\qvec} = \sin(2\varphi_{\kvec,\qvec})$, 
where $\varphi_{\mathbf{k,q}}$ is the angle between $\kvec$ and $\qvec$ 
(see, e.g.,  Eqs. (2.3) in Ref. [\onlinecite{garst}]). 
The CO CM has instead a finite characteristic wavevector $\qvec_c$ and couples to the 
fermion quasiparticle via a weakly momentum dependent coupling $g_{c}(\kvec,\qvec)\approx g_c$ 
(i.e., $d^c_{\kvec,\qvec}\approx 1$).

We assume that these CMs are near an instability and their propagators take the standard Gaussian form, valid 
within a Landau-Wilson approach, and already adopted for models of fermion quasiparticles coupled to 
nearly critical charge [\onlinecite{CDG}] and spin [\onlinecite{mmp},\onlinecite{chubukov}] CMs in cuprates. 
As customary in quantum critical phenomena, different damping processes may lead to different dynamical 
critical exponents $z$, relating the divergent correlation length $\xi$ and time scale $\tau\propto \xi^z$. In the 
case of the nematic instability a multiscale criticality occurs due to the different dynamics of 
transverse and longitudinal fluctuations [\onlinecite{garst},\onlinecite{zacharias}]. 
The longitudinal fluctuations are Landau-overdamped, decay in particle-hole pairs acquiring a dynamical 
exponent $z_\ell=3$, and their propagator is
\begin{equation}
{D}_{\ell}(\gr{q},\omega_n)= -\frac{1}{m_\ell+c_\ell|\qvec|^2+|\omega_n|/|\qvec| + 
\omega^2_n/ \Omega_\ell},
\label{eq:CML}
\end{equation} 
where $\omega_n$ is a boson Matsubara frequency and wavevectors $\qvec$ are henceforth assumed
dimensionless and measured in units of inverse lattice spacing $a^{-1}$ 
(when needed, conventional units are restored in our formulas by replacing $\qvec$ with $a \qvec$). 
Apart from the term $\propto \omega_n^2$, this propagator 
is the same as that in  Eq. (2.14) of Ref. \onlinecite{garst}. 
Transverse fluctuations have instead $z_t=2$, and their propagator is [see, e.g., Eq. (2.15) in Ref. \onlinecite{garst}] 
\begin{equation}
{D}_{t}(\gr{q},\omega_n)= -\frac{1}{m_t+c_t|\qvec|^2+|\omega_n|+\omega_n^2/(\Omega_t|\qvec|^2)}.
\label{eq:CMT}
\end{equation} 
Both propagators, in the static limit ($\omega_n=0$), are peaked at $\gr{q}=0$.
Similarly, the nearly critical CO CM has a dynamical critical index $z_c=2$, with
propagator [see, e.g.,  Eq. (1) in Ref. \onlinecite{enss} or Eq. (2) in Ref. \onlinecite{CDMPHLEKAG}]
\begin{equation}
{D}_{c}(\gr{q},\omega_n)= -\frac{1}{m_c+c_c|\qvec-\qvec_c|^2+|\omega_n|+ 
\omega^2_n/\Omega_c},
\label{eq:CMC}
\end{equation}
peaked at a finite wavevector $\qvec_c$ (actually, at the whole star of equivalent wavevectors). 
This circumstance allows to reabsorb a factor $|\qvec_c|^2$ in the definition of the parameter 
$\Omega_c$, and marks the difference with respect to the propagator of the transverse NCMs, Eq. 
(\ref{eq:CMT}). In the doping regime we are considering, $\qvec_c$ is directed
along the diagonals of the BZ in LSCO with $x<0.05$ [\onlinecite{yam98}]. According to the discussion
in Sec.\,\ref{intro}, we consider instead that CO is absent in YBCO with $p\approx 0.015$.

In Eqs. (\ref{eq:CML}-\ref{eq:CMC}), the parameters $c_{\lambda}$ set the curvature at the bottom 
of the CM dispersions, whereas the parameters $\Omega_\lambda$ set high-frequency cutoffs. The low-frequency 
scales $m_{\lambda}$ are proportional to the inverse squared correlation lengths $\xi_\lambda^{-2}$, thus
being the relevant parameters that measure the distance from criticality. 

\begin{figure}
\includegraphics[width=1.0\linewidth]{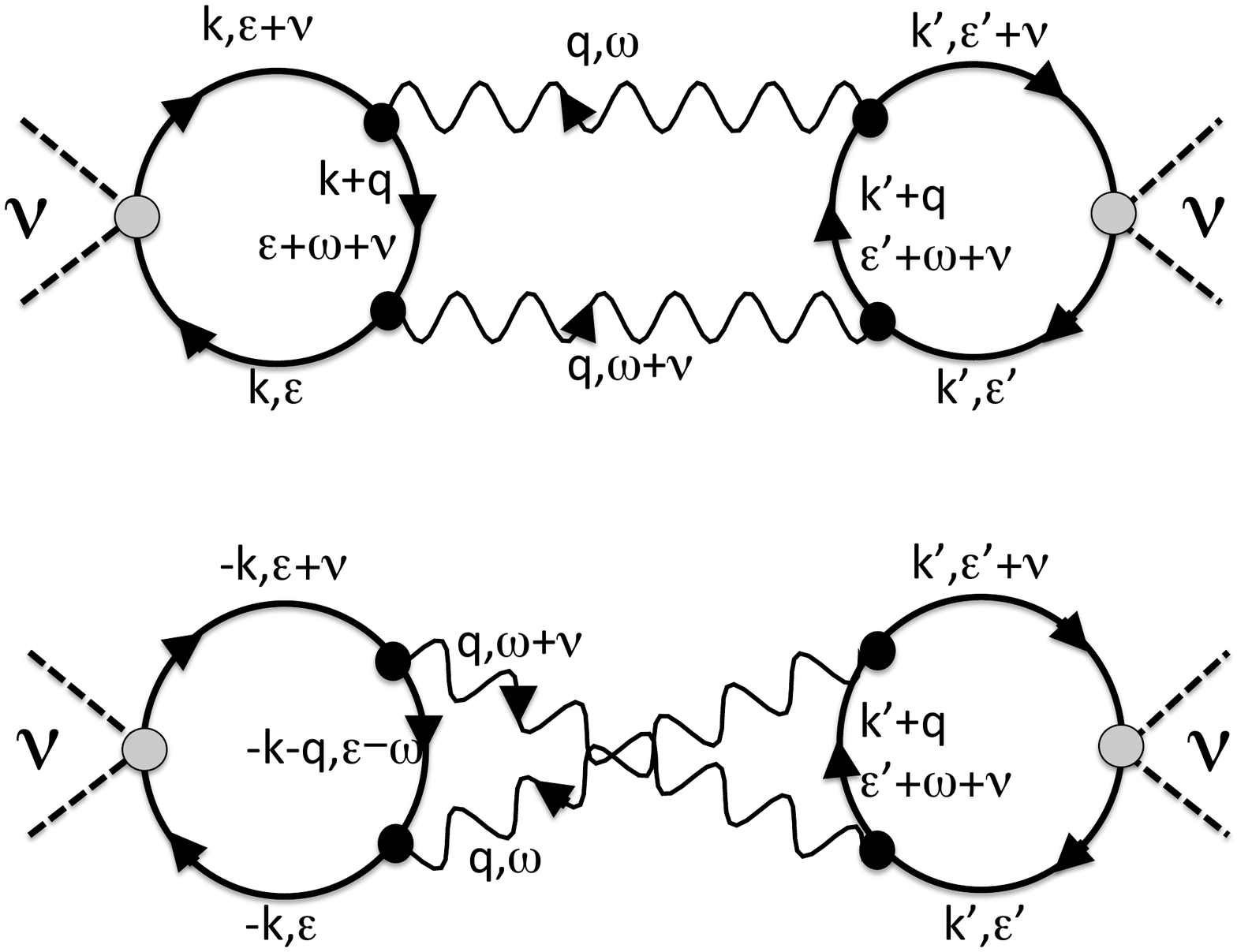}
\caption{Diagrammatic representation of the Raman response due to the excitation of two CMs.
The grey dots represent the $\gamma_{B1g}$ or $\gamma_{B2g}$ form factors. The solid lines are the
propagators of the fermion quasiparticles in the fermionic loops, the wavy lines represent the 
NCM or CO CM propagators [Eq.s (\ref{eq:CML},\ref{eq:CMT},\ref{eq:CMC})], which are coupled to the 
quasiparticles by the coupling functions $g_\lambda(\kvec,\qvec)$ (solid dots).}
\label{raman-diagrams}
\end{figure}

\subsection{The fermionic loop in the Raman response}
\label{floop}

Our theoretical analysis is based on the calculation of the Raman response represented by the Feynman diagrams 
of Fig.\,\ref{raman-diagrams} (more details are given in Appendix \ref{detail}). 
The first step is to calculate the sum of the fermionic loops with attached 
direct and crossed boson lines (see top and bottom diagrams in Fig.\,\ref{raman-diagrams})
\begin{eqnarray}
&&\Lambda_i^{\lambda\eta}(\qvec,\nu_l,\omega_m)=
T\sum_{\mathbf{k},n}\gamma_i(\mathbf{k})g_\lambda(\kvec,\qvec)g_\eta(\kvec,-\qvec)
\nonumber\\
&&\times\left[
G(\mathbf{k}+\mathbf{q},\epsilon_{n}-\omega_{m})+ G(\mathbf{k}+\mathbf{q},\epsilon_{n}+\omega_{m}+\nu_l)    
\right]\nonumber\\
&&\times 
G(\mathbf{k},\epsilon_{n})G(\mathbf{k},\epsilon_{n}+\nu_l),\label{fermloop}
\end{eqnarray}
where $T$ is the temperature, $i=B_{1g},B_{2g}$ labels the form factors, 
$\gamma_{B1g}(\kvec)=\cos(k_y)-\cos(k_x)$ and $\gamma_{B2g}(\kvec)=\sin(k_x)\sin(k_y)$ 
[\onlinecite{Devereaux:2007}], $\nu_l$ is the external Matsubara frequency which, once analytically 
continued, represents the frequency shift between the incoming and the scattered photons, $\omega_{m}$ is 
the Matsubara frequency of one of the boson propagators in Fig.\,\ref{raman-diagrams}
(the other carries $\omega_m+\nu_l$), $\epsilon_{n}$ is the fermion frequency to be summed over in the fermionic 
loop, and $G(\mathbf{k},\epsilon_{n})=(i\epsilon_{n}-\xi_{\mathbf{k}})^{-1}$ is the fermion quasiparticle 
propagator. In Eq.\,(\ref{fermloop}) we exploited the parity of $G(\mathbf{k},\epsilon_{n})$, 
$\gamma_i(\mathbf{k})$, and $g_\lambda(\kvec,\qvec)g_\eta(\kvec,-\qvec)$ with respect to $\mathbf{k}$.

The dependence of the loop on the CM indexes $\lambda$ and $\eta$ is diagonal: the CO CM cannot mix with
the NCMs, having a finite characteristic wavevector, and the $\ell$ and $t$ NCMs cannot mix, because
the product of $g_\ell(\kvec,\qvec)$ and $g_t(\kvec,-\qvec)$, each depending only on the angle between 
$\kvec$ and $\qvec$ and having a different parity, averages to zero when summed with respect to
$\kvec$. This fact entails a selection rule stating that the two NCMs attached to the same fermionic loop must 
be either longitudinal or transverse. The average over the Fermi surface of two couplings with the same 
NCM yields a result that is weakly dependent on $\qvec$ and can be safely approximated to
a constant that can be reabsorbed in the definition of the dimensional coupling $g_\lambda$. Thus 
$\Lambda_i^{\lambda\eta}(\qvec,\nu_l,\omega_m)\equiv g_\lambda^2\delta_{\lambda\eta}\Lambda_i(\qvec,\nu_l,\omega_m)$.

Summing over the fermion frequencies, one obtains the general expression
\[
\Lambda_i(\qvec,\nu_l,\omega_m)=
2\sum_\kvec\frac{\gamma_i(\mathbf{k})\Delta f_{\bf k}
\left[\Delta\xi_{\bf k}^{2}-\omega_{m}(\omega_{m}+\nu_l)\right]}{(\Delta\xi_{\bf k}^{2}+
\omega_{m}^{2})[\Delta\xi_{\bf k}^{2}+(\omega_{m}+\nu_l)^2]},
\]
where $\Delta f_{\bf k}\equiv f(\xi_{\mathbf{k}+\mathbf{q}})-f(\xi_{\mathbf{k}})$,
$\Delta\xi_{\bf k}\equiv\xi_{\mathbf{k}+\mathbf{q}}-\xi_{\mathbf{k}}$, and  $f(z)\equiv
(\mathrm e^{z/T}+1)^{-1}$ is the Fermi function.

The next steps are different in the case of NCMs (with characteristic wavevectors $\qvec\approx 0$)
and of CO CMs (with finite characteristic wavevectors $\qvec_c$), and will be dealt with in Sec.\,\ref{NCMloop}
and Sec.\,\ref{CCMloop}, respectively.
 
\subsubsection{The fermionic loop for NCMs}
\label{NCMloop}

To proceed with the calculation of the fermionic loop in the case of NCMs, we consider that 
the main features of the boson propagators (\ref{eq:CML}) and (\ref{eq:CMT}) are their poles at small momenta
$q=|\qvec|$ and even smaller frequencies, because of their dynamics with
$z_\ell=3$ ($\omega\sim q^3$) or $z_t=2$ 
($\omega\sim q^2$). Thus, expanding the above result for 
small frequencies and keeping the lowest order in the Matsubara frequencies
$\omega_{m}$ and $\omega_m+\nu_l$, one obtains
\[
\Lambda_i(\qvec,\nu_l,\omega_m)
\approx 2\sum_\kvec\frac{\gamma_i(\mathbf{k})\Delta f_{\bf k}}{\Delta\xi_{\bf k}^{2}}\approx 
2\sum_{\kvec}\frac{\gamma_i(\mathbf{k})}{\Delta\xi_{\bf k}}
\frac{\partial f(\xi_{\mathbf{k}})}{\partial\xi_{\mathbf{k}}}.
\]
The summation on $\kvec$ can be transformed into a two-dimensional integral, yielding
\be
\Lambda_i(\qvec)\approx\frac{2M}{\left(2\pi\right)^{2}}\int\int {\mathrm d}k\, 
{\mathrm d}\theta\,\delta(k-k_{F})\frac{\gamma_i(k,\theta)}{\Delta\xi_{\bf k}},
\label{fermiloopncm}
\ee
where $\theta$ is the angle between the wavevector $\kvec$ and the $x$ axis in reciprocal space, and
$M$ is the quasiparticle effective mass. By noticing that the form factor $\gamma_i(\kvec)$, 
calculated on the Fermi surface, depends weakly on $k=|\kvec|$ while it substantially depends 
on $\theta$, one can write $\gamma_{B1g}(k,\theta) \approx \cos(2\theta)\equiv \gamma_{B1g}(\theta)$
and $\gamma_{B2g}(k,\theta) \approx \sin(2\theta)\equiv \gamma_{B2g}(\theta)$. When expanding 
$\Delta\xi_{\bf k}$ one has to keep track of the inverse band curvature $M$ (otherwise the integral vanishes). The
limit $|\qvec|\to 0$ can then be taken, and the final result is that the fermionic loop depends only 
on the angle $\phi$ between $\mathbf{q}$ and the $x$ axis. This dependence can be made explicit
observing that the denominator $\Delta\xi_{\bf k}$ in Eq. (\ref{fermiloopncm}) depends on the cosine of the angle $\theta-\phi$ between
$\kvec$ and $\qvec$. Shifting the variable $\theta-\phi\to\theta$, one is left with $\gamma_i(\theta+\phi)$
in the numerator. Expanding, one has $\gamma_{B1g}(\theta+\phi)=\gamma_{B1g}(\theta)\gamma_{B1g}(\phi)-\gamma_{B2g}(\theta)
\gamma_{B2g}(\phi)$ and $\gamma_{B2g}(\theta+\phi)=\gamma_{B2g}(\theta)\gamma_{B1g}(\phi)+\gamma_{B1g}(\theta)
\gamma_{B2g}(\phi)$. The integral with respect to $\theta$ of the terms with $\gamma_{B2g}(\theta)$ vanishes by symmetry. 
Thus, we finally obtain
\begin{equation}
\Lambda_i(\phi)\approx\frac{M^{2}}{\pi k_{F}^{2}}\gamma_i(\phi).
\label{fermloopresult}
\end{equation}
This result, which is crucial in our development, implies that the original form factor
$\gamma_i(\theta)$ coupling the fermion quasiparticles to the incoming and outgoing photons in 
the Raman vertex, in the integrated form of the loops, is translated into a direct coupling
of the photons to the NCMs with the same form factor $\gamma_i(\phi)$.

\subsubsection{The fermionic loop for the CO CMs}
\label{CCMloop}

The fermionic loop for the CO CMs has been calculated in Ref.\,[\onlinecite{suppa}], and we recall here 
the main results. The main difference with respect to the calculation of Sec.\,\ref{NCMloop}, is that the
propagator (\ref{eq:CMC}) is peaked at finite wavevectors $\qvec_c$. Then, the sum over $\kvec$ in 
Eq. (\ref{fermloop}) is now dominated by the neighborhood of the points along the Fermi surface where 
$\xi_\kvec=\xi_{\kvec+\qvec_c}$, i.e., the so-called hot spots. The results is
\begin{equation}
\Lambda_i(\qvec_c)\approx
\frac{1}{2 \pi^2} \ln \left|\frac{W_+}{W_-}\right| \sum_{HS} \frac{\gamma_{i,HS}}{v_{HS}^2\sin\alpha_{HS}}
\label{fermloopco}
\end{equation}
where $W_\pm$ are the upper and lower cutoffs for the linearized band dispersion at the hot spot, while 
$\gamma_{i,HS}$ and $v_{HS}$ are, respectively, the Raman form factor and the Fermi velocity at $\kvec=\kvec_{HS}$, and
$\alpha_{HS}$ is the angle between the Fermi velocities at the two hot spots connected by the given $\qvec_c$.
For $\qvec_c$ along high symmetry directions (i.e., the axes and the diagonals) of the BZ, the moduli of the 
Fermi velocities in $\kvec_{HS}$ and 
$\kvec_{HS}+\qvec_c$ are equal. As pointed out in Ref.\,[\onlinecite{suppa}], summing over $\kvec$ at {\it fixed} 
$\qvec_c$, various different hot spots are visited, where, due to the above-mentioned symmetry, the form factors 
can have pairwise equal magnitude and equal or opposite signs. As a consequence, the terms in the above hot-spot 
summation can add or cancel each other. This 
induces a ``selection rule'' which, in the case pertinent to the strongly underdoped LSCO, where
$\qvec_c$ is short and directed along the $(\pm 1,\pm 1)$ directions, leads to finite $B_{2g}$ vertex loops, while 
the $B_{1g}$ vertex loops vanish by symmetry.

\subsection{The Raman response}
\label{Rresp}
 
Few considerations are now in order. First of all, the NCM propagators, Eqs. (\ref{eq:CML}) and (\ref{eq:CMT}),
do not depend on the angle $\phi$ and therefore the product of the two fermionic loops entering the diagrams of 
Fig.\,\ref{raman-diagrams} only introduces a multiplicative constant factor, which can be enclosed in the overall 
intensity of the Raman response. However, we emphasize that the $\phi$ integration, to be performed when the summation 
over $\qvec$ is carried out, introduces an important selection rule: The fermionic loops with attached Raman vertices, 
enter pairwise in the response diagrams of Fig.\,\ref{raman-diagrams} and must both be of the same symmetry, $B_{1g}$ or 
$B_{2g}$. Similarly, the CO propagator (\ref{eq:CMC}) depends only on the magnitude of the deviation
of $\qvec$ from $\qvec_c$. In this case the Raman response is given by a first summation on all the possible 
$\qvec_c$ of $\left[\Lambda_i(\qvec_c) \right]^2$ and an internal integral over $|\qvec -\qvec_c |$ of two CO CM 
propagators.  

Thus, both for the two nematic CMs and for the CO CM, the sum of the two diagrams of Fig.\,\ref{raman-diagrams} reads
\[
\chi_{i,\lambda}(\nu_l)=K_{i,\lambda}T\sum_{n}\int_0^{\bar q} {\mathrm d}q\,\frac{q}{\bar q^2}
D_\lambda(q,\omega_{n})D_\lambda(q,\omega_{n}+\nu_l),
\]
where $\bar q\sim 1$ is the momentum cutoff, its precise value being re-absorbable in a multiplicative rescaling 
of the parameters of the CM propagator, $q=|\qvec|$ for the NCMs, and $q=|\qvec-\qvec_c|$ for the CO CMs.
The factor $K_{i,\lambda}$ comes from the product of two fermionic loops 
and is proportional to $g_\lambda^4$ (each loop $\Lambda_i$ carrying two fermion-CM coupling constants). 
For NCMs, $K_{i,\lambda}\equiv M^4 g_\lambda^4{\langle [\gamma_i(\phi)]^2\rangle}/
\left(\pi k_F^2\right)^2$, with $\lambda=\ell$ or $t$, and
$\langle [\gamma_i(\phi)]^2\rangle$ is the angular average of the square of the function
$\gamma_i(\phi)$ that appears in Eq. (\ref{fermloopresult}), whereas for the CO CM
we have [cf. Eq. (\ref{fermloopco})] $K_{i,c}=g_c^4\sum_{\qvec_c} [\Lambda_{i}(\qvec_c)]^2$, that
vanishes in the $B_{1g}$ ($B_{2g}$) channel for diagonal (vertical/horizontal) $\qvec_c$.
This ``selection rule'' is the only place where the finite wavevector of the CO CM 
plays a role within our nearly critical theory of Raman absorption. 
This selection rule is instead absent in the case of the NCMs, which are peaked at $\qvec=0$.

The analytic continuation to real frequencies $i\nu_l\rightarrow\omega+i\delta$ and the use 
of the spectral representation of the boson propagators finally yield the Raman response
\bea
\chi_{i,\lambda}^{\prime\prime}(\omega)&=& A_{i,\lambda}\int_{-\infty}^{+\infty}
{\mathrm d}z\, \left[b(z_{-})-b(z_{+})\right]  
\int_{0}^{1} {\mathrm d}q\, \nonumber \\
&\times&  2q\,F_\lambda(z_+,q)F_\lambda(z_-,q),
\label{longabs}
\eea
where $b(z)\equiv(\mathrm e^{z/T}-1)^{-1}$ is the Bose function, and we performed 
the customary symmetrization $z\rightarrow z-\frac{\omega}{2}\equiv z_{-}$ and
$z+\omega\rightarrow z+\frac{\omega}{2}\equiv z_{+}$, which makes explicit the fact that
Eq.\,(\ref{longabs}) is an odd function of $\omega$. The constant multiplicative prefactors, including
those transforming the Raman susceptibility into the measured Raman response, are reabsorbed 
in the parameters $A_{i,\lambda}$. 
Unfortunately a fully analytical expression for $A_{i,\lambda}$ cannot be given, the 
Raman response being affected by resonance effects that prevent even order-of-magnitude estimates. However, 
whenever we studied the contribution of critical CMs in situations where the prefactors can be explicitly calculated, 
like optical conductivity [\onlinecite{CDMPHLEKAG},\onlinecite{enss}], or angle resolved photoemission spectra 
[\onlinecite{mazza}], we always found that the dimensionless coupling constants are of order one, in a regime 
of moderate coupling.

The spectral density of the longitudinal NCMs is
\[
F_\ell(z,q)=\frac{\frac{z}{q}}{\left(m_\ell+c_\ell q^2-\frac{z^{2}}{\Omega_\ell}\right)^{2}
+\frac{z^{2}}{q^2}},
\]
while the spectral density of the transverse NCMs is
\[
F_t(z,q)=\frac{z}{\left(m_t+c_t q^2-\frac{z^2}{\Omega_t q^2}\right)^{2}+ z^{2}}.
\]
Finally, for the CO CMs we find
\[
F_c(z,q)=\frac{z}{\left(m_c+c_c q^2-\frac{z^2}{\Omega_c}\right)^{2}+ z^{2}}.
\]

\begin{figure}
\includegraphics[width=1.\linewidth]{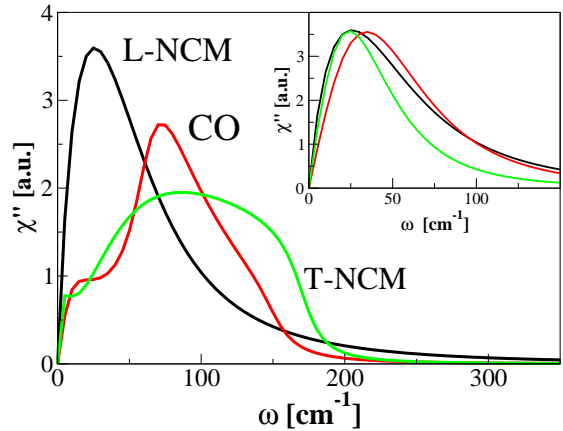}
\caption{Schematic representation of the AL-like Raman response of the three different CMs: longitudinal L-NCM 
(black curve), transverse (T) NCM (green curve), and CO CM (red curve). Inset: AL-like Raman response from different 
CMs, but taken with the same nearly critical set of parameters ($m_\lambda=17$ cm$^{-1}$, $\Omega_\lambda=70$ cm$^{-1}$,
$c_\lambda=3.16$) The amplitudes are instead rescaled by factors of order one to bring all responses to a common maximal height
for easier comparison. The coloring of the lines is the same as in the main panel. 
}
\label{sketch}
\end{figure}

The anomalous peak of the Raman response, both in LSCO and YBCO, is strongly temperature dependent, 
it shrinks and softens upon reducing $T$. This behavior is naturally encoded in the temperature dependence of the
mass $m_\lambda$ of the CMs.
In general, the low-frequency scale $m_\lambda$ controls the slope of the Raman response, while the scales
$\omega_1\sim \sqrt{m_\lambda\Omega_\lambda}$ and $\omega_2\sim \sqrt{(m_\lambda+c_\lambda)\Omega_\lambda}$ set the frequency 
window over which the spectral function of the corresponding CM is sizable.
However, the different dynamical properties and values of the parameters of the CMs mirror 
into different shapes of the AL-like Raman responses, which are schematically represented in Fig.\,\ref{sketch}.
We point out that the curves displayed in this 
panel do not exhaust all the possible regimes of parameters, and only represent the corresponding CM in the regime
where, after a thorough analysis, they were found to better reproduce the various features of the Raman response.
The inset of the same figure reports instead the behavior of Raman absorption spectra (from the AL 
processes) due to the various CMs.  In this inset, while we rescale the height to bring all responses to the same maximal height, 
we use the same nearly critical set
of parameters to highlight the differences arising purely from the different form of the propagators and
dynamical critical index $z$. Apparently the shape of the spectra is quite similar, but the behavior upon changing 
the mass is different 
on a quantitative level. In particular we found that the $z=2$ propagators shift the position of the maxima upon reducing $m$
more rapidly than the NCM $z=3$ propagator.
Since we apply a strict fitting protocol (see below in Section\,\ref{results}), which fixes all parameters and follows the 
temperature evolutions of the main peaks by only changing $m$, these different behavior affects in a substantial way the 
accuracy of the fits. An inspection of
Fig. 3 in Ref. \onlinecite{suppa} shows that the fits with a $z=2$ CO-CM are not very accurate at low temperatures. 
Instead we will see in the
next Section that the $z=3$ NCM does a much better job within the adopted strict fitting protocol and therefore
it will be considered henceforth as the primary (i.e., most critical) CM.
The additional shoulder in the spectra of LSCO, 
is instead better reproduced by the CO curve in Fig.\,\ref{sketch} than by the broader T-NCM curve, when both CMs are taken
in the regime of parameters apt to describe this spectral feature. Therefore 
at these doping levels the CO CM acts as the secondary CM in LSCO.

\section{Results}
\label{results}

\subsection{Raman absorption in Y$_{0.97}$Ca$_{0.03}$Ba$_2$Cu$_3$O$_{6.05}$}

\begin{figure}
\includegraphics[width=1.\linewidth]{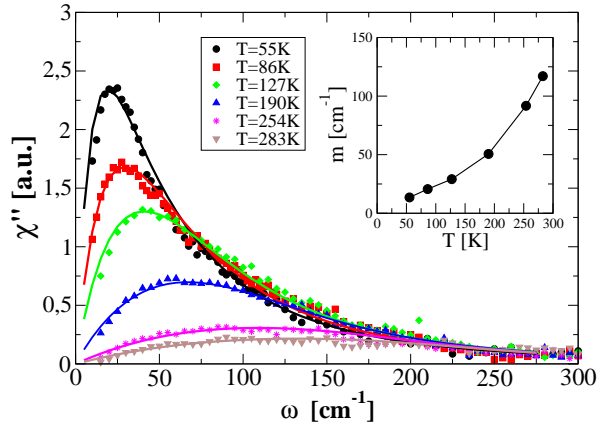}
\caption{Subtracted experimental Raman absorption spectra in the $B_{2g}$ channel, at various temperatures, for 
YBCO at $p\approx 0.015$ (symbols). The theoretical fits (solid lines) consider the contribution of the longitudinal NCM 
only. The fitting parameters are  $c_\ell=0.63$ cm$^{-1}$, $\Omega_\ell=110$\,cm$^{-1}$, $A_\ell=5.0$ (a.u.). 
The inset reports 
the temperature dependence of the mass of the longitudinal NCM (black circles).}
\label{ybco}
\end{figure}

An anomalous Raman absorption at low frequencies, up to few hundreds of cm$^{-1}$, is experimentally found 
in the $B_{2g}$ channel in lightly doped YBCO with $p\le 0.05$ [\onlinecite{tassini2008}]. Since, however, the 
whole spectra also display broad absorptions up to electronic energy scales, we first extract the specific anomalous 
low-frequency contributions. To this purpose, we subtract from the low-temperature spectra the spectra 
obtained at the highest measured temperature. This subtraction is delicate because at temperatures below
about 150-200\,K, the spectra are characterized by the formation of a pseudogap over a frequency range of several
hundreds of cm$^{-1}$, which reduces the electronic background. Then, the simple
subtraction leads to regions of negative absorptions, which are obviously meaningless. In the 
Appendix B we provide the detailed procedure adopted to circumvent this drawback. In Fig.\,\ref{ybco}, the data
processed according to the previous procedure, are shown for $p\approx 0.015$.

The experimental lineshape clearly resembles the L-NCM spectrum in Fig.\,\ref{sketch}, which is narrow due to the 
$z_\ell=3$ 
damped dynamics of the corresponding CM, whose temperature dependence is ruled by the mass $m_\ell$. Indeed, 
the data in Fig.\,\ref{ybco} are best fitted with the only contribution of longitudinal NCMs. In the spirit of our 
nearly-critical approach, we only adjust their mass $m_\ell(T)$, while keeping all other parameters (i.e., the
high-frequency cutoffs of the CM propagator, the $c_\ell$ coefficients, and the overall intensity coefficient $A_\ell$) 
fixed at all temperatures. This strict procedure was already successfully adopted in Ref.\,[\onlinecite{suppa}] and seems 
to us the most suitable to pinpoint the quantum nearly-critical character of the collective excitations responsible 
for the anomalous Raman absorption. The fits with this restricted procedure turn out to be quite good. Of course they 
could be further improved if this constrained procedure were relaxed. The fits reproduce well the lineshapes and the 
strong temperature dependence of the peak, encoded in the rapid decrease of the mass with temperature, as shown in the 
inset of Fig.\,\ref{ybco}. From this inset, it is evident that $m_\ell(T)$ decreases with $T$. Its linear extrapolation 
starting from high temperature should vanish at some finite critical temperature for the onset of nematicity 
($\approx 125$\,K), if static order would occur. However, at lower temperatures, the mass seems instead to saturate, 
likely indicating that nematic order stays short-ranged and dynamic.

\begin{figure}
\includegraphics[width=1.\linewidth]{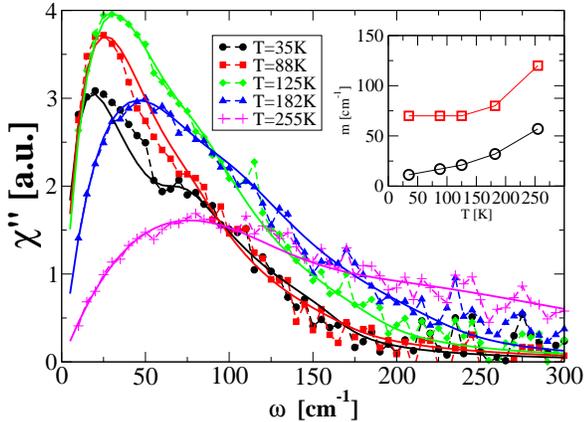}
\caption{Subtracted experimental Raman absorption spectra in the $B_{2g}$ channel, at various temperatures, for LSCO 
at $x=0.02$ (symbols). The theoretical fits (solid lines) consider the contribution of the longitudinal NCM and of  
the CO CM, with $c_\ell=3.16$ cm$^{-1}$, $c_c=333$ cm$^{-1}$, $\Omega_\ell=70$ cm$^{-1}$, $A_\ell=8.3$ (a.u.). The other 
fitting parameters are reported in Fig.\,\ref{param}. The inset reports the temperature dependence of the mass of the 
longitudinal NCM (black circles) and of the CO CM (red squares).}
\label{fig2-fit002}
\end{figure}

\subsection{Raman absorption in La$_{2-x}$Sr$_x$CuO$_4$}
\begin{figure}
\includegraphics[width=1.\linewidth]{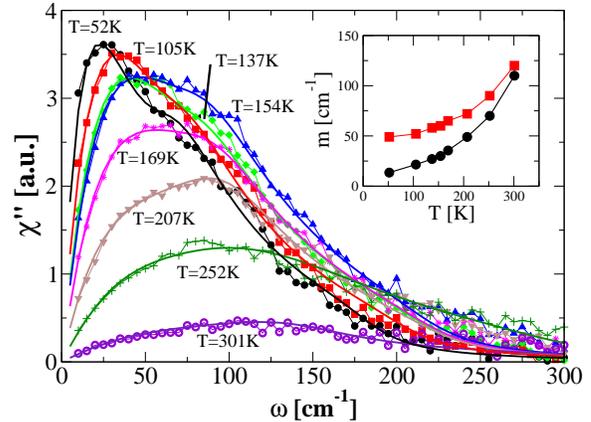}
\caption{Subtracted experimental Raman absorption spectra in the $B_{2g}$ channel, at various temperatures, for LSCO 
at $x=0.04$ (symbols). The theoretical fits (solid lines) consider the contribution of the longitudinal NCM and of 
the CO CM, with $c_\ell=3.16$ cm$^{-1}$, $c_c=333$ cm$^{-1}$, $\Omega_\ell=50$ cm$^{-1}$. $A_\ell=7.14$ (a.u.). The other
fitting parameters are reported in Fig.\,\ref{param}. The inset reports the temperature dependence of the mass of the 
longitudinal NCM (black circles) and of the CO CM (red squares).}
\label{fig3-fit004}
\end{figure}

Figs.\,\ref{fig2-fit002} and \ref{fig3-fit004} report the experimental Raman spectra in the $B_{2g}$ channel, for 
LSCO samples at doping $x=0.02$ and $x=0.04$ and various temperatures. The raw data were again processed according 
to the procedure described in the Appendix B. As mentioned above, the anomalous Raman absorption observed in LSCO 
is characterized by a lineshape that is more complex than in YBCO, and displays a peculiar shoulder or, at low $T$, 
even a secondary peak, see Fig.\,\ref{fig3-fit004}. The anomalous peak and the shoulder (or secondary peak) both depend 
on temperature, but their frequency and intensity are not simply related by constant multiplicative factors. The 
shoulder (or secondary peak) becomes stronger with increasing doping. This indicates that the excitations responsible 
for this absorption have a distinct dynamics.

Again these absorptions are described by the AL-like processes (direct and crossed, see Fig.\,\ref{raman-diagrams}). 
Owing to the selection rules found in Sec.\,\ref{qpcm}, the response due to two (or more) CMs is the sum of the 
responses associated with each individual CM. As already mentioned, our thorough analysis showed that the primary 
anomalous absorption should be attributed to the longitudinal NCM, which has the stronger dynamical behavior. Within 
our context, the transverse NCM and the CO CM are the two candidates for the shoulder (or secondary peak). Looking at 
the lineshape of the two CMs reported in Fig.\,\ref{sketch}, it is easy to convince oneself that the best choice for 
a good fit is the CO CM, due to its much more pronounced peaked form at intermediate frequency. We also attempted a 
fit with the transverse NCM. At $x=0.02$ we obtained a reasonable fit taking a very large and almost temperature 
independent CM mass, which is hardly compatible with our assumption of nearly critical CMs. Moreover, at $x=0.04$, 
when the shoulder evolves into a secondary peak, the attempt failed completely. Thus, we ruled out a contribution 
of transverse NCMs.

\begin{figure}
\includegraphics[width=.9\linewidth]{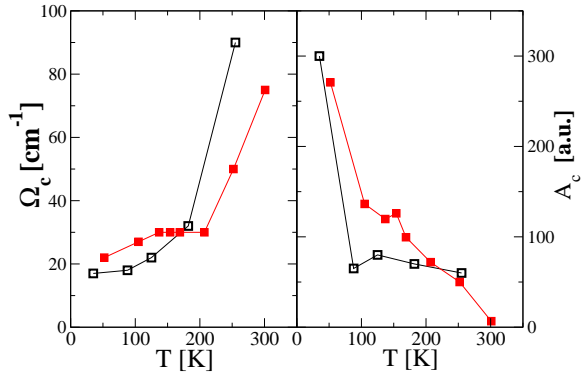}
\caption{ (a) High-frequency cutoff ${\Omega_{c}}$ for the CO CM for a sample at $x=0.02$ (black empty squares) 
and at $x=0.04$ (red filled squares). (b) Amplitude coefficients $A_c$ for the CO CM
for a sample at $x=0.02$ (black empty squares) and at $x=0.04$ (red filled squares). }
\label{param}
\end{figure}

Again, having attributed the main peak to the more critical longitudinal NCM, we describe the low-frequency side 
of the spectra by only adjusting the mass $m_\ell(T)$ of this excitation, while keeping all other parameters of this 
mode (i.e., the high-frequency cutoff of the CM propgator, the $c_\ell$ coefficient, and the overall intensity coefficient 
$A_\ell$) fixed at all temperatures, within the temperature range considered here. Thus, we obtain the marked temperature 
dependence of the longitudinal NCM mass, which is reported in the insets of Figs.\,\ref{fig2-fit002} and 
\ref{fig3-fit004}. On the other hand, the complete quantitative agreement between data and theoretical fits is only 
obtainable by adjusting more freely the secondary CO CM. This mode is therefore allowed to vary its parameters with $T$, 
as reported in Fig.\,\ref{param}. The temperature dependence of the CO CM parameters 
$\Omega_c$ and $A_c\propto g_c^4$ likely reflects an increasing damping and a decreasing coupling to the fermion 
quasiparticles with increasing $T$. Of course, the estimates and variations of these parameters may be 
quantitatively affected if the 
constraint of $T$-independent parameters for the longitudinal NCM (but for its mass $m_\ell$) were relaxed. 
Furthermore, we cannot exclude that static nematic order has eventually occurred, e.g., in the sample with $x=0.02$
at the lowest temperature. In this case our analysis, which is only valid above the critical temperature, 
should be modified to deal with a broken-symmetry phase. This might reflect in a reduction 
of the primary peak, due to the freezing of NCM fluctuations, and could be the cause of the non monotonic behavior of 
the peak height as a function of $T$, observed in the sample with $x=0.02$. To asses the 
occurrence of static nematic order at low temperature, a systematic experimental investigation in this temperature 
regime is needed.

\section{Discussion and Conclusions}
\label{disconcl}
Our analysis showed that the anomalous Raman absorption observed in underdoped cuprates can be interpreted 
in terms of direct excitation of nearly critical CMs (see Fig.\,\ref{raman-diagrams}). The strong temperature 
dependence of the mass (i.e., inverse square correlation length) of the ``primary'' CM, identified as the longitudinal 
NCM (with dynamical critical index, $z_\ell=3$), captures the correspondingly strong variation of the spectra. This CM 
alone fully accounts for the spectra of YBCO. In LSCO, instead, a distinct ``secondary'' CM, with different 
dynamical critical index $z=2$, is needed to reproduce the composite lineshape. Within the two candidates 
considered in our scheme (transverse NCM and CO CM), our fits indicate that the CO CM is the most suitable. 

\begin{figure}
\includegraphics[width=1.\linewidth]{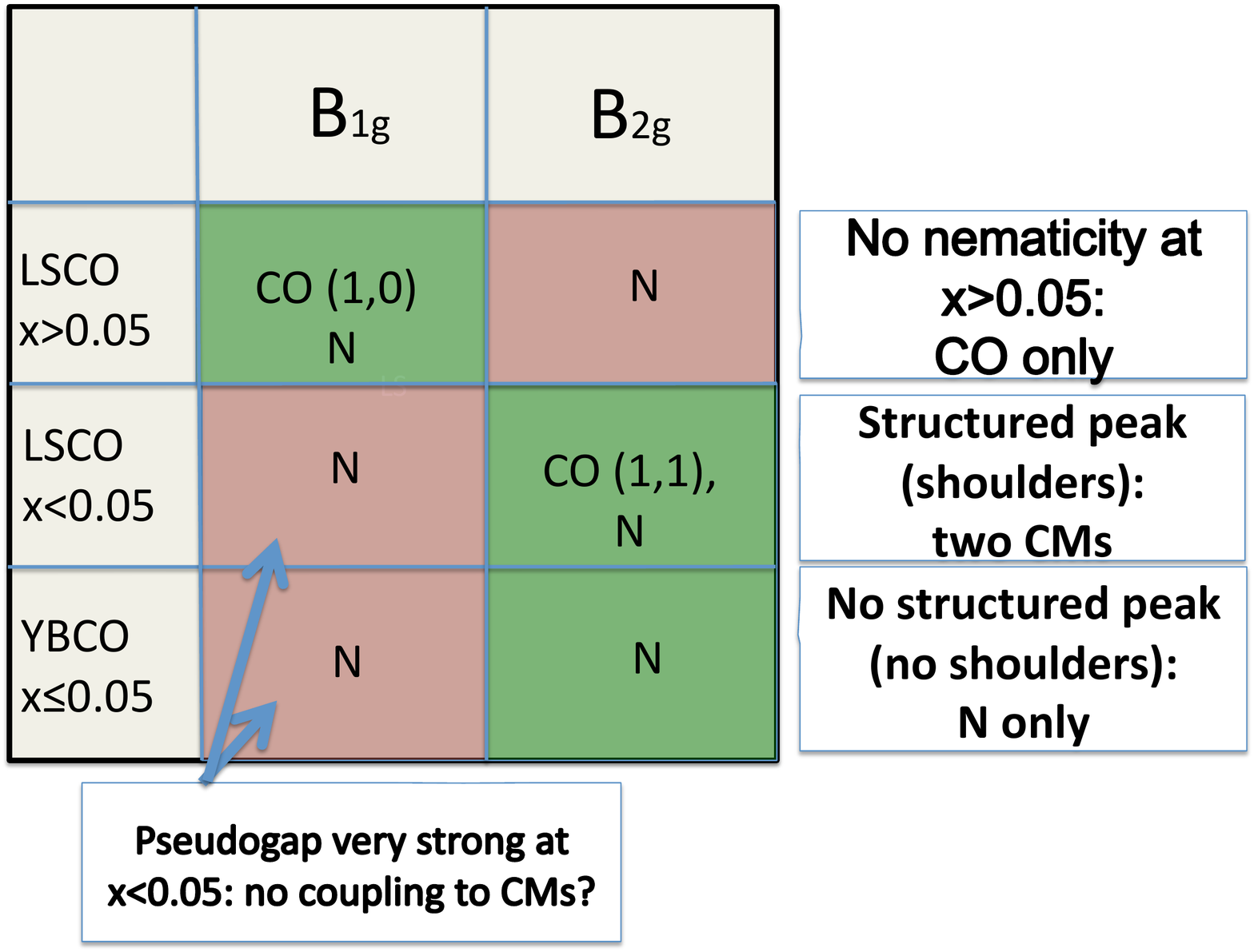}
\caption{Schematic comparison of the theoretical expectations and the experimental observation of an anomalous Raman 
absorption. The theoretically involved CMs are indicated with N in the case of the NCM, while for CO we also report the 
direction of the characteristic wavevector, as established by inelastic neutron scattering. The related symbols only 
appear in the box where they are expected to contribute on the basis of symmetry arguments. The experimental 
observation of an anomalous Raman absorption is depicted as a green case in the column of the corresponding channel. 
Red cases indicate, instead the lack of anomalous Raman absorption in experiments. Our remarks and possible 
indications (in boldface) are contained in the comment boxes.}
\label{schema}
\end{figure}

For symmetry reasons the secondary CO CM cannot occur in all channels: The first two rows in the sketch of 
Fig.\,\ref{schema} summarize the findings of Ref.\,\onlinecite{suppa} in LSCO as far as CO is concerned:
The correct CO (i.e., with finite $\qvec_c$ in the direction compatible with inelastic neutron scattering 
experiments) appears as an observed absorption (green case) {\it only} in the theoretically predicted channel.

Two questions still remain to be answered, in order to complete the scheme of Fig.\,\ref{schema}. First of all, the NCMs
would equally contribute to the $B_{1g}$ and $B_{2g}$ channels. Therefore, they would not only add to the CO fluctuations 
that give absorption in the $B_{1g}$ channel at larger doping $(x>0.05)$ in LSCO, but would also give rise to 
absorption in the $B_{2g}$ channel. Since this is not observed (the corresponding box is red in Fig.\,\ref{schema}), 
we infer that NCMs disappear in LSCO at $x>0.05$ (see the comment box in the first row of Fig.\,\ref{schema}). This 
is consistent with the observation of an increasingly stronger stripe order at higher doping [\onlinecite{suppa}], 
where CO CM alone [along the (1,0) and (0,1) directions of the BZ] accounted for the anomalous Raman absorption 
at $x=0.10$ and $0.12$.

The second related question is: If the NCMs are present and contribute to the absorption in $B_{2g}$ at low doping both 
in LSCO and YBCO, why are they not visible in the (for them allowed) $B_{1g}$ channel? As yet, we do not have a 
definite answer. We argue that the strong pseudogap occurring in lightly doped cuprates at $T< 200$\,K could play a 
key role in suppressing the $B_{1g}$ absorption. Specifically, the $B_{1g}$ form factors select the quasiparticles in 
the fermionic loops of Fig.\,\ref{raman-diagrams} precisely from the BZ regions where the pseudogap is largest. 
Therefore, only the quasiparticles in the remaining Fermi arcs, mostly weighted by the $B_{2g}$ form factors, remain 
to couple the Raman photons with the NCMs. In Appendix C, we obtained a numerical estimate of this suppression, 
finding indeed that it can be substantial.

\begin{figure}
\includegraphics[width=1.\linewidth]{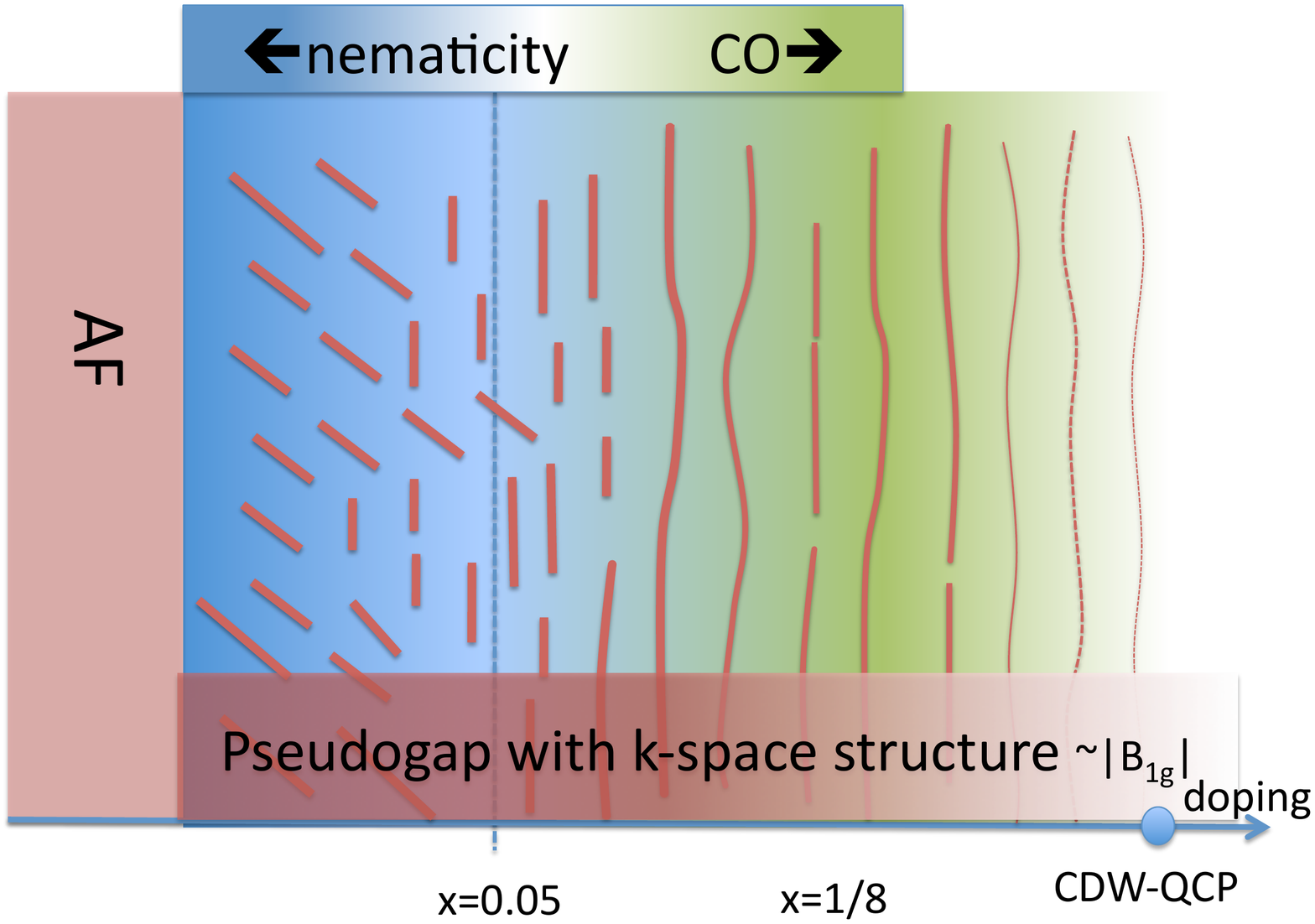}
\caption{Schematic evolution of the nematic (blue region) and stripe CO (green region) phases in underdoped 
cuprates with doping (disregarding superconductivity). The pseudogap region, where the Raman response in the $B_{1g}$ 
channel is expected to be suppressed, is highlighted in red. Upon increasing doping, the nematic phase evolves into a 
CO phase, which in turn vanishes at a CO quantum critical point around optimal doping. The orientation of the segments 
and/or stripes may change with cuprate family and doping.}
\label{phase-diagram}
\end{figure}

Based on the above discussion, we can draw the schematic ``phase diagram'' for underdoped cuprates reported in 
Fig.\,\ref{phase-diagram}. Despite its speculative character, it is compatible with various theoretical and 
experimental findings, and accounts for the assessed relevance of nematic order in cuprates 
[\onlinecite{hinkov08,haug,daou,mesaros2011,lawler2010}]. It also complies with the proposal of a nematic order 
resulting from the melting of stripes [\onlinecite{KFE}] or of a nematic or smectic phase 
in strongly underdoped LSCO (and possibly YBCO) [\onlinecite{capati}], arising from the aggregation of doped 
charges in short segments (blue region). The orientation of these segments breaks the lattice $C_4$ 
rotational symmetry preparing the route to CO at higher doping, when the segments merge into stripes 
(green region). The fluctuating character of the nematic phase should give rise to nearly critical fluctuations of 
the form of Eqs. (\ref{eq:CML}) and (\ref{eq:CMT}). On the other hand, CO fluctuations become prominent by 
increasing doping and appear in the $B_{1g}$ channel above $x=0.05$. In LSCO these fluctuations are present (in 
the diagonal directions of the BZ) also at $x<0.05$ and contribute to the $B_{2g}$ absorption, but 
the tendency of CO fluctuations to become more relevant at larger doping is clearly visible by comparing 
Figs.\,\ref{fig2-fit002} and \ref{fig3-fit004}. At the same time, the insets of Figs.\,\ref{fig2-fit002} and 
\ref{fig3-fit004} also display an increase of the low-temperature limit of the correlation length of the CO CM
upon increasing doping. Hence, nematic and CO fluctuations coexist in very underdoped LSCO, the predominance shifting 
from nematic to CO with increasing doping. This indicates a continuous evolution from the nematic (charge segment) 
phase to the stripe phase where charge and spin degrees of freedom are tightly bound, yielding a 
definite relation between spin and charge incommensurabilities (typical of the stripe phase). We relied on this 
relation to implement our symmetry-based selection rules for LSCO. On the other hand, our finding that NCMs alone 
are relevant in YBCO at very low doping supports the idea that oriented charge segments may occur in this material 
as well, accounting for the order-parameter-like disappearance of the incommensurability in the 
spin response with increasing temperature [\onlinecite{SCDGL},\onlinecite{capati}], as observed in Refs. 
[\onlinecite{hinkov08},\onlinecite{haug}]. The lack of CO fluctuations at low doping and the opposite doping dependence 
of the charge and spin characteristic wavevectors [\onlinecite{blancocanosa}] indicate a nematic-to-CO switching 
different from that in LSCO. However, both materials seem to eventually evolve into a charge-density-wave phase 
ending into a quantum critical point around optimal doping, as theoretically proposed [\onlinecite{CDG}] and 
recently observed [\onlinecite{blancocanosa}].

\acknowledgments 
S.C. and  M.G. acknowledge financial support form the University of Rome ``Sapienza'' with the Project Awards n. 
C26H13KZS9.

\appendix
\section{Details on the calculation of the Raman response}
\label{detail} 

Inspection of Fig.\,\ref{raman-diagrams} shows that the diagrammatic structure of the Raman response due to
the excitation of two CMs involves two inequivalent fermionic loops (on the left of the
two diagrams), which multiply two CM propagators and the fermionic loop on the right 
of the diagrams. Calling $\Lambda_i^{\lambda\eta}(\qvec,\nu_l,\omega_m)$ the sum of the 
two different fermionic loops (frequencies and momenta are those displayed in Fig.\,\ref{raman-diagrams}),
we can write the expression for the sum of the two diagrams as
\begin{eqnarray*}
\chi_{ij}^{\lambda\eta}(\nu_l)&=&T\sum_{\qvec,m}
\Lambda_i^{\lambda\eta}(\qvec,\nu_l,\omega_m)\nonumber\\&\times& D_\lambda(\qvec,\omega_m)
D_\eta(\qvec,\omega_m+\nu_l)\,L_j^{\lambda\eta}(\qvec,\nu_l,\omega_m),\nonumber
\end{eqnarray*}
where $i,j=B_{1g},B_{2g}$, $\lambda,\eta=\ell,t,c$, and
$L_j^{\lambda\eta}(\qvec,\nu_l,\omega_m)$ stands for the fermionic loop in the right 
part of the diagrams. The above expression can be made symmetric also with respect to the latter fermionic 
loop, observing that the integrated $\kvec'$ can be changed into $-\kvec'$, and the fermionic Matsubara
frequency $\epsilon_n'$ can be shifted to $\epsilon_n'-\nu_l$
(frequencies and momenta are those displayed in the fermionic loop on the right of both diagrams in 
Fig.\,\ref{raman-diagrams}). Then, exploiting the parity of the
CM propagators with respect to both momentum and frequency arguments, we can take $\qvec\to -\qvec$
and $\omega_m\to -\omega_m$, $\nu_l\to -\nu_l$. Summing the two equivalent expressions and
dividing by 2 we are finally led to calculate 
\begin{eqnarray*}
\chi_{ij}^{\lambda\eta}(\nu_l)&=&\frac{T}{2}\sum_{\qvec,m}
\Lambda_i^{\lambda\eta}(\qvec,\nu_l,\omega_m)\nonumber\\&\times& D_\lambda(\qvec,\omega_m)
D_\eta(\qvec,\omega_m+\nu_l)\,\Lambda_j^{\lambda\eta}(\qvec,\nu_l,\omega_m).\nonumber
\end{eqnarray*}
This expression has the formal structure of a Raman response where to CMs are directly excited by the
scattered electromagnetic radiation, and $\Lambda_i^{\lambda\eta}(\qvec,\nu_l,\omega_m)$ plays the
role of an effective Raman vertex, resulting from the sum of the fermionic loops with attached 
direct and crossed boson lines in Fig.\,\ref{raman-diagrams}. In Sec. \ref{floop} we discuss the calculation 
of the effective vertex $\Lambda_i^{\lambda\eta}(\qvec,\nu_l,\omega_m)$. This calculation is further specialized to
the cases of NCMs and CO CMs in Secs. \ref{NCMloop} and \ref{CCMloop}, respectively.

\section{Fitting procedure of the anomalous Raman absorption}
\begin{figure}
\includegraphics[width=1\linewidth]{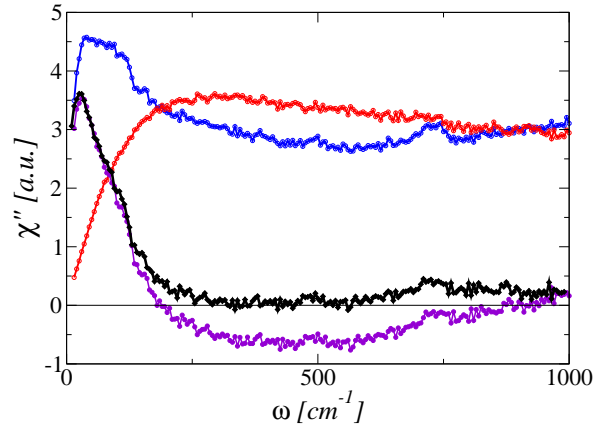}
\caption{Subtraction procedure on the raw Raman data of a LSCO sample at $x=0.04$ and $T=52$\,K (blue curve and 
symbols). The red curve and symbols correspond to the raw data at $T=331$\,K. Once the latter are subtracted from 
the former, the purple curve and symbols are obtained. To eliminate the negative absorption, the parabola
$\chi^{\prime\prime}_{b}=\omega (1000-\omega)/500^2$, with $\omega$ in cm$^{-1}$, is added to finally yield the 
absorption reported with the black curve and symbols.}
\label{subtract}
\end{figure}

To fit the anomalous contribution of the Raman absorption due to two virtual CMs, as represented in 
Fig.\,\ref{raman-diagrams}, one needs to subtract the regular part of the spectra arising, e.g., from the dressed 
quasiparticles. However, the subtraction procedure has to face the problem of pseudogap formation occurring in the 
underdoped regime: At substantially high temperatures (above 300\,K) there is no pseudogap, which 
instead sets in below 200\,K. The anomalous absorption peak we are interested in starts to appear on 
top of the (pseudogapped) broad absorption spectra at lower $T$. Thus, when the non-pseudogapped spectra at 
$T\approx 300$\,K are subtracted from the low-temperature pseudogapped ones, a negative differential absorption is 
found over the frequency range of the pseudogap. Although this is not crucial for the qualitative description 
of the anomalous peaks, to get quantitatively more precise fits we exploit the fact that the pseudogap sets in 
rather rapidly and, once established, depends only very weakly on $T$. Therefore, we add a smooth parabolic contribution 
$\chi^{\prime\prime}_{b}=\omega (\Omega_{MAX}-\omega )/[B(T)]^2$, with $\omega$ in cm$^{-1}$ and $T$ in K, just 
designed to cancel the negative part at each temperature. For YBCO we take $\Omega_{MAX}=1000$, $B(55)=760$, 
$B(86)=860$, $B(127)=1100$, while at $T=190,\,254,\,282$\,K no compensation is needed because the pseudogap is not 
open. For LSCO at $x=0.02$ we take $\Omega_{MAX}=800$, $B(35)=500$, $B(88)=500$, $B(125)=600$, $B(182)=1000$, 
while at $T=255$\,K again no compensation is needed because the pseudogap is not open. The same procedure is 
carried out at $x=0.04$, with $\Omega_{MAX}=1000$, $B(52)=600$, $B(105)=600$, $B(137)=800$, $B(154)=1000$, 
$B(169)=800$, $B(207)=1500$, $B(252)=2000$, $B(301)=2000$. Fig.\,\ref{subtract} exemplifies the procedure for LSCO
with $x=0.04$ at $T=52$\,K. The blue curve represents the raw data, from which we subtract the red data at 
$T=331$\,K, obtaining the purple curve with unphysical negative absorption. The pseudogap effect is then corrected 
by the addition of the smooth parabolic contribution, leading to the final black curve. Once these differential 
spectra are thus brought to have a zero background we proceed to fit the strongly $T$-dependent anomalous peaks.

\section{Pseudogap, Fermi arcs, and Raman response suppression}
The strongly underdoped phase of cuprates is characterized by the presence of a pseudogap that strongly suppresses 
the low-energy electronic degrees of freedom. In particular, the electronic states in the so-called antinodal 
regions of the BZ, around $(\pm \pi,0),(0,\pm \pi)$, are gapped, while the so-called nodal states, along the 
$(\pm 1,\pm 1)$ directions, survive giving rise to Fermi arcs which shrink upon lowering temperature and doping. 
In this appendix we investigate the effects of this suppression of the low-energy electronic states on the 
coupling between the Raman vertices and the NCMs. Indeed the fermionic loops entering the diagrams of 
Fig.\,\ref{raman-diagrams} involve the integration over fermionic degrees of freedom coupled to the nearly critical 
CMs, with the low-energy fermions being the most effective in coupling to the low-energy CMs. Therefore the opening 
of gaps in the electronic spectra naturally entails a substantial reduction of the overall response of the CM. 
However, the Raman vertices $\gamma_i(\kvec)$ weight differently the fermionic states along the Fermi surface and it 
is therefore quite natural that the fermionic loops are differently suppressed depending on the channel. 
To estimate this effect is the aim of this appendix. More specifically, we will consider the NCMs only, because the
CO modes in the very underdoped LSCO were shown in Ref.\,[\onlinecite{suppa}] to be only visible in the
$B_{2g}$ channel.  So it would be meaningless to compare the pseudogap effects in the two Raman channels.
On the contrary, the NCM are singular at $\qvec\approx 0$ and therefore should give a strong contribution to the 
Raman response in both channels. This appendix will instead demonstrate that the interplay of Raman vertices and 
momentum dependence of the pseudogap strongly suppress the loop in the $B_{1g}$ channel in comparison to the 
$B_{2g}$ case.

\begin{figure}
\includegraphics[width=1.\linewidth]{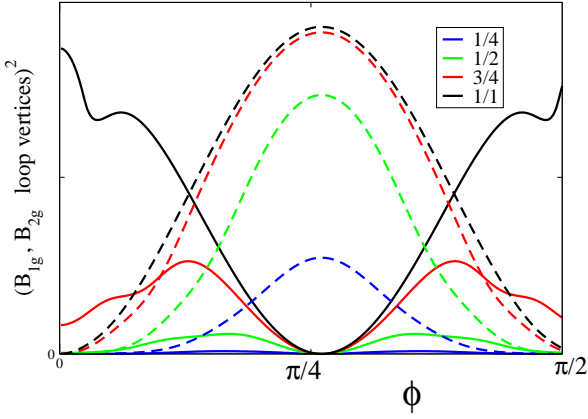}
\caption{Square of the fermionic loops as a function of the angle $\phi$ calculated according to Eq. (\ref{loop}) in 
the $B_{1g}$ channel (solid lines) and in the $B_{2g}$ channel (dashed lines). The $F(\theta)$ function [Eq.(\ref{Ftheta})] has been set 
to produce arcs shrinking the Fermi surface by a factor 1 ($\Theta_M=\pi/4$, whole Fermi surface, black curves), 
0.75 ($\Theta_M=3\pi/16$, red curves), 
0.5 ($\Theta_M=\pi/8$, green curves), and 0.25 ($\Theta_M=\pi/16$, blue curves). The parameter $\Delta_\theta^2=0.05$,
was set to smoothen the angular cutoff producing the arcs. 
The boson momentum $q$ has been chosen such that $q^2/2k_F^2=0.01$, 
while $M=1$.}
\label{b1gb2garcs}
\end{figure}

We adopt the simplifying assumption of a circular Fermi surface and we start from the 
expression for the fermionic loop, Eq. (\ref{fermiloopncm}). Since the NCMs are singular
at small $\qvec$, we expand in this limit the quantity $\Delta\xi_{\bf k}\equiv \xi_{\kvec}-\xi_{\kvec+\qvec}$.
Expanding up to order $q^2$ the denominator and exploiting the $\delta$ function to perform the
integral along the radial momentum variable, one obtains
\be
\Lambda_i \approx \frac{2M^2}{(2\pi)^2} \int_0^{2\pi}{\mathrm d}\theta F(\theta)
\frac{\gamma_i(\theta)}{1-\frac{q^2}{2k_F^2}+\cos(2\theta-2\phi)},
\label{loop}
\ee
where $\phi$ is the angle between $\qvec$ and the $x$ axis in reciprocal space. At this stage we have 
phenomenologically introduced a function 
\be 
\label{Ftheta}
F(\theta)=\sum_{n}\frac{1}{1+e^{\{[\theta-(2n-1)\pi/4]^2-\Theta_{M}^2\}/\Delta_\theta^2}},
\ee 
with $n=1,2,3,4$, which simulates the effect of the pseudogap on the Fermi surface of the $1-4$ quadrants. 
Specifically this function leaves the states near the diagonal untouched, while 
for $\Theta_{M}<\pi/4$ it rather sharply suppresses 
the integration in the gapped antinodal regions for $\theta$'s far from the nodal direction $\theta=\pi/4$ (for 
$\Theta_M=\pi/4$ one recovers the full ungapped Fermi surface). This essentially restricts the integration in Eq. 
(\ref{loop}) to the angles of a Fermi arc allowing to explore the different action of the Fermi surface shrinking 
on the value of the fermionic loop. The parameter $\Delta_\theta$ measures how rapidly the pseudogap 
is switched on and off along the FS, and we take it to be much smaller than $\pi/4$.

Fig.\,\ref{b1gb2garcs} displays the square of the fermionic loops in the two Raman channels as a function of the angle 
$\phi$ between the boson transferred momentum $\qvec$ and the $x$ axis. The calculation clearly shows the increasingly 
strong suppression of the $B_{1g}$ fermionic loop (solid curves) upon reducing the length of the Fermi arcs.
The suppression is much less  pronounced in the $B_{2g}$ fermionic loop (dashed lines). These results are rather natural 
because the pseudogap suppresses the states that more effectively contribute to the $B_{1g}$ loop, while the 
fermion quasiparticles contributing more to the $B_{2g}$ channel survive in the Fermi arcs.

\bibliographystyle{prsty_no_etal}

\end{document}